# Topics in Quantum Dynamics and Coherence for Quantum Information Processing

**Vladimir Privman**

Department of Physics, Clarkson University, Potsdam, NY 13699, USA
Web page: www.clarkson.edu/Privman

We outline selected trends and results in theoretical modeling of quantum systems in support of the developing research field of quantum information processing. The resulting modeling tools have been applied to semiconductor materials and nanostructures that show promise for implementation of coherent, controlled quantum dynamics at the level of registers of several quantum bits (qubits), such as spins. Many-body field-theoretical techniques have been utilized to address a spectrum of diverse research topics. Specifically, the theory of decoherence and more generally the origin and effects of quantum noise and the loss of entanglement in quantum dynamics of qubits and several-qubit registers has been advanced. Qubit coupling mechanisms via the indirect exchange interaction have been investigated, and quantum computing designs have been evaluated for scalability. We outline general and specific research challenges, the solution of which will advance the field of modeling "open quantum systems" to further our understanding of how environmental influences affect quantum coherence and its loss during quantum dynamics.

---

# 1. Introduction

We survey and outline selected results in the new research filed of "controlled quantum dynamics," aimed at investigating general aspects of quantum noise, as well as single- and multi-qubit decoherence, robustness of entanglement, and novel schemes for two-qubit interactions, mediated via the qubits' coupling with a many-body bath of modes (e.g., acoustic phonons, conductions electrons). Study of "open quantum systems," with new challenges suggested by the emerging field of quantum information, requires utilization and development of new field-theoretical many-body techniques for the description of quantum dynamics. These studies have facilitated evaluation of scalability of existing and emerging quantum computing schemes, with, in our case, emphasis on experimentally explored spin- and quantum-dot based quantum computer designs.

Both creation of entanglement, by induced indirect-exchange Ruderman-Kittel-Kasuya-Yosida (RKKY) type interactions, and its loss due to quantum noise have been quantified by developing calculation techniques for the system's dynamics, with the environmental (bath) modes traced out, to describe the induced interaction and quantum noise in a unified treatment. Several open problems of general as well as practical interest, the latter motivated by recent experimental developments, will be discussed.

The reported research has been focused on the control/coherence of quantum dynamics via/due to qubit-qubit interactions and on quantum noise, mediated/caused by coupling to external baths, in particular, acoustic phonons and electrons in conduction channels connecting quantum-dot and impurity-atom *spin qubits* in semiconductor-material heterostructures. The latter, RKKY-type mechanism of spin-spin (qubit-qubit) interaction provides a unique opportunity to incorporate scalable control of entanglement into solid state quantum computing schemes. Calculations of interactions and decoherence have been carried out for systems of relevance to the ongoing measurement and interaction experiments, referenced in the following sections. In Section 2, we generally mention and cite our modeling results. In Section 3, we review the present status of research work in the field of evaluation of decoherence. In Section 4, we describe new research challenges in the topics involving induced interactions (RKKY-type) and their interplay with quantum noise. Finally, in Section 5 we address some longer-term technical, as well as conceptual research challenges, in the physics of open quantum systems as applied to modeling qubits subject to environmental noise and at the same time to quantum control, for quantum information processing.

## 2. Overview of Results

It has been widely recognized that decoherence (quantum phase noise) [1] and, for several qubits, the associated loss of entanglement, are the main obstacles to scalable quantum computation. Part of an evaluation for any quantum computing scheme should be identification of system parameter ranges for which the level of quantum noise is within the conditions for fault-tolerant quantum error correction. Even if large-scale quantum computation were to be regarded as a futuristic concept, coherent quantum control of systems and structures on the nanoscale, but larger than single atoms and molecules, has now become realistic in experiments that control quantum dynamics well beyond the traditional energy-level spectroscopy. Reduction of quantum noise to maintain quantum coherence over the experimental time scales is key to probing processes and concepts that were until recently out of reach of experimental observations, including such amazing recent discoveries as observation of an electron spin resonance (ESR) signal from a single spin [2,3], or probing RKKY-type indirect exchange interaction that, via conduction electrons, influences the dynamics of just two spins (of electrons located in separate quantum dots) [4].

Here we outline our broad program of research to advance theoretical evaluation of decoherence, entanglement, and interactions for a variety of quantum computing systems. From our group's original focus, in the late nineties, on the quantum-Hall-effect quantum computing scheme [5-7], our studies have broadened to rather general results on decoherence [1,8-10] and, more recently, disentanglement and induced RKKY-type interactions for two qubits [11-13], as well as to additive measures of decoherence for multi-qubit registers [1,14]. Specific model calculations have been carried out for a range of spin- and quantum-dot-based quantum computing schemes in semiconductor nanostructures, e.g., [1,10,15,16]. While the focus has been on semiconductor materials, most of these general results actually apply to any qubit-based quantum computing scheme.

## 3. Evaluation of Decoherence during Quantum Gate Functions

Let us briefly summarize the present status of the field of evaluation of decoherence. It has been recognized that for times, $t$, larger than $\hbar/kT$, which is about $2.5 \times 10^{-14}$ sec at room temperatures of 300K, Markovian approximation schemes, e.g., [17,18], can be used. Traditionally, these have been utilized for primarily single-qubit calculations of the relaxation and dephasing time scales, $T_1$ and $T_2$, and of the asymptotic large-time properties of the density matrix elements of the qubit, with the environmental (bath) modes traced out.

Quantum information processing necessitates the development of techniques applicable for the "short-time" regime $t < \hbar/kT$, because most quantum-computation schemes are for systems at very low temperatures, for which gate times are shorter than or comparable to $\hbar/kT$. For such short times, the bath mode correlations with the system of qubits cannot be entirely treated within the no-memory, immediate-rethermalization Markovian approximations. We have developed a new calculational approach, which *does not include* the Markovian assumption, designed specifically for this regime, and reported applications, e.g., [8-10,16] for several semiconductor quantum computing schemes.

Presently, the level of quantum noise in the idling state, for a single qubit, as well as during the simplest quantum gates, has been well quantified [1,10,19-38]. The calculation involves evaluation of the qubit's density matrix, followed by calculation of numerical measures of the degree of decoherence. The calculations can be done within both our short-time and the Markovian approximations, and the largest (the worst-case scenario) noise measure values can then be used to test the system for scalability as candidate for quantum computer designs.

We have introduced new quantitative measures of decoherence which are approximately *additive* [1,10,14] as long as the noise for each qubit is small (for the short duration of a quantum gate function), and therefore these measures can be used to further extending the results to *several-qubit quantum registers*, in terms of the noise measures of the individual qubits, even when these qubits are *strongly entangled*.

In these calculations, the Hamiltonian of the qubit system and bath of environmental modes, is modeled by

$$H = H_S(t) + H_B + H_{SB} \ ,$$

(1)

where

$$H_S(t) = H_i + H_g(t) \ .$$

(2)

Here $H_i$ is the Hamiltonian of the idling qubit register, while $H_g(t)$ represents time-dependent gate control. The Hamiltonian of the environment is given by $H_B$, and $H_{SB}$ is its interaction with the qubits. The environment is often modeled as a bath of bosonic quantum fields (e.g., phonons in solid state).

There have been few available results for evaluation of quantum noise *during general time-dependent* gate functions. The reason is that the time dependence introduced by non-zero $H_g(t)$, requires time-ordering in the evaluation of the qubit evolution operator, which makes many standard techniques for dynamical calculations inapplicable. Recently, we initiated a new approach [39] based on a variant of the Magnus expansion studied in quantum chemistry applications, and we have obtained the first results for a gate-controlled qubit interacting with a boson bath of modes.

Recently reported first experimental successes in coherent control of single electron spins in quantum dots, e.g., [40], pose an interesting theoretical challenge to develop systematic calculational techniques for general-time-dependence gate-controlled qubits, as well as attempt to establish additivity criteria to enable evaluation of the level of noise for registers of several such controlled qubits. This suggests that some interesting work can be carried out in the future, developing such evaluation schemes and their applications for spin and other quantum computing architectures. In summary, in this section we outlined trends in evaluation of single-qubit decoherence, as well as mentioned studies of additivity of decoherence measures for the most direct extension of quantum-noise level estimates to several-qubit registers.

## 4. Indirect Exchange Interaction and Quantum Noise

Modeling of an open quantum system within the Hamiltonian description, Eqs. (1)-(2), involves significant technical complications. Indeed, while the system Hamiltonian can be just single- or two-spin, the bath-mode Hamiltonian involves many modes, in the simplest case a collection of noninteracting bosonic fields,

$$H_B = \sum_k \omega_k b_k^\dagger b_k \ .$$

(3)

The qubit-bath interaction can involve terms of the form

$$H_{SB} = \Lambda_S \sum_k g_k b_k^\dagger + \Lambda_S^\dagger \sum_k g_k^* b_k + ... \ ,$$

(4)

where $\Lambda_S$ is a system operator, as well as more complicated expressions in the case of fermionic bath fields (conduction electrons). There can usually be several interaction terms involved in Eq. (4), and in Eq. (3), if more than one qubit is studied, and for various bosonic polarizations, for instance for acoustic phonons, or for more than one bath dominating the relaxation of the system.

However, and even more importantly, there are *physical assumptions* that have to be invoked to supplement this description. The open-quantum system dynamics cannot be entirely specified by the Hamiltonian for the system and the selected bath of external modes. We have to also model the effects of the thermalization imposed by the "rest of the universe" on the bath (as well as consider the choice of the initial conditions for the whole system — an issue that represents an important challenge on its own, which will not be addressed here). These topics have a long history in the nuclear magnetic resonance (NMR) and ESR literature, primarily for single, idling spins treated within Markovian approximations, e.g., [17,18,41-44].

Recent experimental efforts to create and maintain entanglement of two electron spins in gate-formed double quantum dots with interaction mediated by electrons in conducting channels [4,45], have focused interest on a new set of challenging problems. The bath modes, while serving as the main source and mediator of the quantum noise causing decoherence, will also cause loss of entanglement of two (and more) spins. However, the same modes, via their interactions with the qubits, can also mediate an indirect exchange (RKKY-type) interaction. The latter can actually create entanglement and, for appropriate initial conditions, drive the approximately coherent quantum dynamics for some time, though ultimately, for long enough times, the quantum noise effects will dominate and cause thermalization (unless the qubits are perturbed by quantum-control gate function external potentials).

Let us remark that the original RKKY calculation corresponds to the zero-temperature evaluation of the coupling induced between localized magnetic moments by a "bath" of conduction electrons. The ideas to utilize various solid-state excitations as a medium to couple qubits for quantum information processing have been advanced by several groups in various settings [5-7,12,13,46-50]. The main advantage of such coupling is that it presumably should allow for larger distances between qubits and therefore for their easier fabrication and control.

Besides the need to properly incorporate the *physics* of the environmental influences in the open-quantum-system description, which is not a fully sorted-out problem to date and is usually handled within phenomenological approaches, one should also aim at *tractability* of the resulting equations for the quantities of interest, for instance, the density matrix of the qubits, $\rho_S$, after the environmental effects were traced out. Typically, a Markovian approximation scheme can lead to a quantum Liouville equation of the type

$$i\hbar\dot{\rho}_S = \left[H_S, \rho_S\right] + (\text{coherent terms}) + (\text{noise terms}) \ . \tag{5}$$

The hope is to be able to derive the "coherent" terms (which could be represented as Hermitian additions to $H_S$ in the commutator) and the noise terms due to the bath modes.

The added terms in Eq. (5), induced by the system's interactions with the bath modes, are usually evaluated within the second-order (in the interaction strength) expansion, as well as other approximations [12,13,18,50] aimed at making the results tractable for calculations. For more than one qubit, the qubit-qubit entanglement is created both by the internal qubit-qubit interactions in $H_S$, and by the induced interactions, which are part of the coherent terms (whereas other coherent terms involve single-qubit operators and represent Lamb shifts).

The noise terms, however, act to unravel the entanglement. Indeed, recently results have been reported for a system of two qubits, suggesting that the noise-induced decoherence of two entangled noninteracting qubits and their disentanglement are closely related [11,51], with the suppression of entanglement taking place at least as fast as the product of the factors that describe the suppression of each qubit's coherence. These results were obtained both for a Markovian-type noise model and for the short-time regime, but limited to uncorrelated noise sources for each qubit.

We have reported the first systematic calculations [12,13,50] of the combined effects of the induced interaction and quantum noise due to the phonon or (interacting, correlated) conduction electron bath, with applications for qubit geometries suggested by possible experiments [2-4,45]. While the actual calculations are quite formidable, the expression for the induced interaction is relatively compact, because it only involves components of two spin operators. As an illustration, here is the result for the calculated [50] indirect exchange interaction between two spins at separation $\mathbf{d}$, mediated by acoustic phonons,

$$H_{\text{int}}(t) = - \sum_{m=x,y,z} \alpha_{n_m}^m \omega_c^{n_m} \sigma_m^{(1)} \sigma_m^{(2)} \frac{2\Gamma(n_m)\,\text{Re}\left(1 + i\,\omega_c\left|\mathbf{d}\right|/c_s\right)^{n_m}}{\left[1 + \left(\omega_c\left|\mathbf{d}\right|/c_s\right)^2\right]^{n_m}} \ . \quad (6)$$

Here $\sigma_{m=x,y,z}^{(1),(2)}$ are the Pauli matrixes of the two qubits, $c_s$ is the speed of sound, and $\omega_c$ is the phonon cut-off frequency. For a system of two spins (of outer electrons) of P-donor impurities, placed in bulk Si at a distance **d** apart, and subject to a constant uniform magnetic field, $H_z$, the exchange interaction is primarily due to the longitudinal acoustic (LA) phonons [50]. The coupling is super-Ohmic: we can assume that $n_m = 3$ in (6) — the small-frequency dependence of the standard Caldeira-Leggett spectral function, and that the coupling parameters due to the spin-orbit interaction, satisfy $\alpha_{n_x}^x = \alpha_{n_y}^y \ll \alpha_{n_z}^z$, with $\alpha_3^z \omega_c^2 \approx 8.4 \cdot 10^{-10}$. The cut-off frequency is due to the localization of the wave function, $\omega_c = c_s / a_B \approx 9.3 \times 10^{12} \mathrm{s}^{-1}$, which in this case is much smaller than the lattice Debye cut-off ($a_B \approx 10^{-9} \mathrm{m}$ is half of the effective Bohr radius of an electron at the P-impurity).

The expression for the noise effects is much more cumbersome [13,50], and model-assumption dependent as is further commented on later. The interplay of the induced interaction and quantum noise, can be used to control the creation of time-dependent entanglement, as illustrated in Figure 1 (see page 19). The obtained results suggest future research into both the buildup of the entanglement and its unraveling, for interacting (and ultimately also controlled, via time-dependent external "gate-function" interactions) qubits subject to quantum noise due to various bosonic and fermionic bath modes, in both the short-time and Markovian regimes. Entanglement should be explored as a fragile resource for quantum information processing, and for the relation [11,51] of disentanglement to decoherence.

The concurrence is by now a generally accepted, calculable in closed form measure of quantum-information content of two-qubit entanglement [52]. For the dynamics of two spins interacting with an acoustic-phonon environment, the concurrence as a measure of entanglement can reveal a rather nontrivial behavior; see Figure 1. We point out, however, that with the full reduced density matrix of the system available, other quantities can be also calculated as needed. e.g., [53].

# 5. Challenges in Many-Body Modeling in Quantum Information Science

In this section we address several longer-term open problems. We believe that, with the actual experimental probes now being carried out at the nanoscale, these problems have become more pressing, and some have been suggested by experimental developments. For definiteness, we consider time-independent $H_S$ from now on.

Perhaps the most fundamental of these problems is the matter of thermalization vs. bath-induced coherent-dynamics effects in open quantum systems. The "textbook" approach to thermalization [17,42-44] has been to assume that, for large enough times, the time evolution of the system plus bath is not just covered by the combined Hamiltonian, but is supplemented by the instantaneous loss-of-memory (Markovian) approximation, which introduces irreversibility and imposes the bath temperature on the reduced system dynamics in the infinite-time limit, which is then approached as the density matrix elements assume their thermal values, according to

$$\rho_S(t) \to e^{-H_S/kT} \Big/ \text{Tr}\Big( e^{-H_S/kT} \Big) \, , \qquad (7)$$

at exponential relaxation (diagonal) and decoherence (off-diagonal) rates defining the time scales $T_{1,2}$.

However, it turns out that the traditional phenomenological no-memory approximations, yielding thermalization, the Fermi golden rule for the transition rates, etc., assume in a way too strong a memory loss: they erase the possible bath contributions to the coherent part of the dynamics at shorter times, such as the Lamb shift for a single system as well as the induced RKKY interactions for a bi-partite system. Indeed, while relaxation leading to Eq. (7) is driven by the "on-shell" exchanges with the bath, it is the memory of (correlation, entanglement with) the bath modes that drives, via virtual exchanges, the induced interaction. Actually, the "on-shell" condition (imposed by the so-called secular approximation, see, e.g., Section 8.1.3 in [17]) also underestimates additional decoherence at short times — the "pure" or "adiabatic" contribution to the off-diagonal dephasing — that has thus been estimated by using other approaches, e.g., [1,8,9,16,33].

Perhaps the simplest way to recognize the ambiguity is to ask if the Hamiltonian in Eq. (7) should have included the bath-induced interaction terms (not shown)? Should the final Boltzmann distribution correspond to the energy levels / basis states of the original "bare" system or the one with the RKKY-interaction / Lamb shifts, and more generally, a bath-renormalized, "dressed" system?

There is presently no consistent treatment that will address in a satisfactory way all the expected *physics* of the bath-mode effects on the dynamics. The issue is partly technical, because we are after a *tractable*, rather than just a "foundational" answer. It is well accepted that the emergence of irreversibility cannot be treated within tractable and calculationally convenient approaches derived directly from the microscopic dynamics: phenomenology has to be appealed to. However, even allowing for phenomenological solutions, all the known tractable schemes yielding the indirect exchange interaction, treat thermalization in a cavalier way, resulting typically in noise terms corresponding to getting the infinite temperature limit at large times, which thus artificially avoids the issue of which $H_S$ should enter in Eq. (7). And, as mentioned, the established schemes that yield thermalization at large times, lose some intermediate- and short-time dynamical effects.

Thus, we have discussed the challenges in formulating unified treatments that will cover all the (or most of the interesting) dynamical effects, covering several time scales, from short to intermediate times (for induced interaction effects and pure decoherence) to large times (for the onset of thermalization), while providing a tractable calculational (usually perturbative, many-body) scheme. This discussion also alludes to several other interesting conceptual challenges in the theory of open quantum systems.

Let us presently comment on the issue of the bath-mode interactions with each other, as well as with impurities, the latter particularly important and experimentally relevant [45,54] for conduction electrons as carriers of the indirect-exchange interaction. Indeed, the traditional treatment of open quantum systems has assumed noninteracting bath modes. When the bath mode interactions had to be accounted for, we have treated the added effects either perturbatively [6], or, for strong interaction, such as Luttinger-liquid electrons in a 1D channel, we took [12] the collective excitations as the new "bath modes." Generally, however, especially in Markovian schemes, one has to seek approaches that do not involve certain double-counting. Indeed, the assumption that the bath modes are at a fixed temperature, could be possibly considered as a partial accounting for the effects of the mode-mode and mode-impurity interactions, because these interactions can contribute to thermalization of the bath, on par with other influences external to the bath. This may be particularly relevant for phonons that always have strong anharmonicity for any real material. In a way this problem fits with the previous one: we are dealing with effects that can be, on one hand, modeled by added terms in the total Hamiltonian but on the other hand, may be also mixed up in the process of thermalization that is modeled by actually departing from the Hamiltonian description and replacing it with Liouville equations that include noise effects. While all this sounds somewhat "foundational," recent experimental advances, interestingly, bring these challenges to the level of application that requires tractable, albeit perhaps phenomenological solutions that can to be directly confronted with experimental data.

There are other interesting topics to be considered, for instance, the question of whether additional sources of quantum noise are possible? It has been recently established [55] that potential difference between two leads (reservoirs, or baths, of electrons) can be a source of quantum noise with the potential difference playing the role of the temperature parameter. What, then, about a system (qubit) coupled to two thermal baths at different temperatures? Will the resulting heat transfer (energy flow between the bath via the system) generate added quantum noise?

As an example of a more practical issue as an open problem, let us mention the possible effect of the sample geometry on phonon and conduction electron induced relaxation and interactions. We have already explored [12] the one-dimensional aspect of the electron gas in a channel. Indeed, electrons are easy to confine by gate potentials. The situation for phonons, however, is less clear: can geometrical effects modify, and particularly reduce, their quantum-noise generation capacity, or change the induced interactions? Our preliminary studies [50] seem to indicate strong overall geometry dependence of the exchange interactions. However, the situation is not entirely clear, especially for the strength of the noise effects, and suggests future explorations because recent experiments with double dot nanostructures in Si membranes [56] indicate that true nanosize confinement (due to the sample dimensions) of otherwise long-wavelength modes (in the transverse sample dimensions) is now possible and will have dramatic effect on the phonon spectrum and, as a result, on those physical phenomena that depend on the phonon interactions with electron spins.

Finally, we point out that several recent experiments, e.g. [57-62], have explored aspects of coherence and control in semiconductor nanostructures for quantum computing. While the progress has been impressive, there is no clear "winner" system. It is likely that the future promising designs will be hybrid, based on the presently explored schemes. It is notable that the ensemble-NMR quantum computing emulation approaches have presently reached partial control and manipulation of 12 qubits [63], which allows consideration of "algorithmic" concepts in multi-qubit system design while the other approaches catch up.

In summary, we have outlined ongoing theoretical research and a selection of challenging open problems to address, in support of the emerging field of quantum information, ultimately aimed at achieving quantum control of multi-qubit systems.

We acknowledge and thank our many co-authors for rewarding scientific collaborations in topics of quantum information: P. Aravind, M.-C. Cheng, A. Dementsov, L. Fedichkin, A. Fedorov, M. L. Glasser, M. Hillery, S. Hotaling, W. B. Johnson, G. Kventsel, R. G. Mani, D. Mozyrsky, V. Narayanamurti, J. A. Nesteroff, Y. Pershin, S. Saikin, M. Shen, D. Solenov, D. Tolkunov, Y.-H. Zhang, and the late I. Vagner.

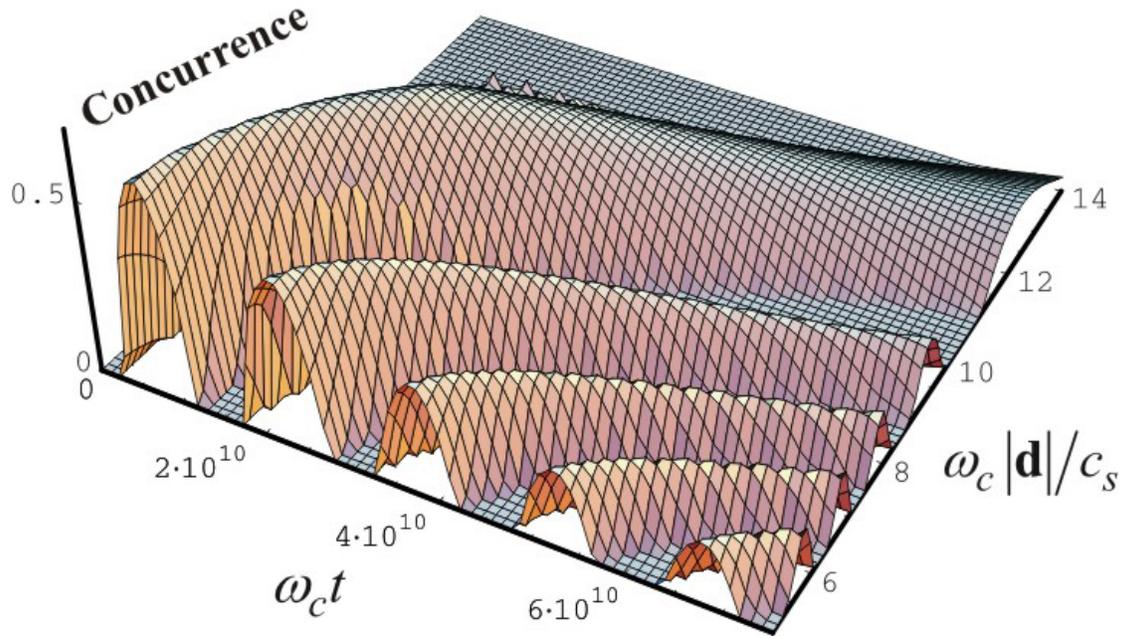

**Figure 1:** The concurrence, drawn as a function of time for various qubit separations, quantifies the entanglement of two spin-1/2 qubits, due to the interaction and quantum noise induced by a bath of acoustic phonons. The parameters here were for P-donor electrons in Si, at $T = 1\,\mathrm{K}$, in external magnetic field $H_z = 3 \times 10^4\,\mathrm{G}$. In terms of the single-spin basis $|\pm\rangle \propto |\uparrow\rangle \pm |\downarrow\rangle$, the two spins were initially in the state $|++\rangle$, whereas the bath was initially thermalized at the temperature $T$.